\documentclass{egpubl}
\usepackage{eurovis2024}

\SpecialIssuePaper         

\CGFccby

\usepackage[T1]{fontenc}
\usepackage{dfadobe}
\usepackage{tabu}                      
\usepackage{booktabs}                  
\usepackage{lipsum}                    
\usepackage{mwe}                       
\usepackage{stackengine}
\usepackage{tabularx}
\usepackage{amsfonts}
\usepackage{amsmath}
\usepackage{bbm}
\usepackage{subfig}
\usepackage[dvipsnames]{xcolor}
\usepackage{colortbl}
\usepackage[normalem]{ulem}
\usepackage{enumitem}

\usepackage{cite}  
\BibtexOrBiblatex
\electronicVersion
\PrintedOrElectronic
\ifpdf \usepackage{graphicx} \pdfcompresslevel=9
\else \usepackage[dvips]{graphicx} \fi

\usepackage{egweblnk}

\title
    {Guided By AI: Navigating Trust, Bias, and Data Exploration in AI-Guided Visual Analytics \vspace{-1em}}

\author[S.\ Ha, S.\ Monadjemi, A.\ Ottley]
{\parbox{\textwidth}{\centering 
        Sunwoo Ha$^{1}$,
        Shayan Monadjemi$^{2}$,
        Alvitta Ottley$^{1}$
        }
        \\
{\parbox{\textwidth}{\centering 
         $^1$ Washington University in St. Louis, St. Louis, MO\\
         $^2$ Oak Ridge National Laboratory, Oak Ridge, TN
       }
}
}

\DeclareMathAlphabet{\mathcal}{OMS}{cmsy}{m}{n}

\newcommand{\ml}[0]{{\textsc{ml}}}
\newcommand{\ai}[0]{{\textsc{ai}}}
\newcommand{\va}[0]{{\textsc{va}}}

\newcommand{\easy}[0]{{\textsc{easy}}}
\newcommand{\hard}[0]{{\textsc{hard}}}
\newcommand{\none}[0]{{\textsc{none}}}
\newcommand{\conf}[0]{{\textsc{conf}}}
\newcommand{\key}[0]{{\textsc{kwd}}}
\newcommand{\keyconf}[0]{{\textsc{kwd+conf}}}
\newcommand{\ctrl}[0]{{\textsc{ctrl}}}
\newcommand{\inv}[0]{{\textsc{investigations}}}
\newcommand{\usage}[0]{{\textsc{usage}}}
\newcommand{\trust}[0]{{\textsc{trust}}}

\definecolor{fig4blue}{HTML}{3241AC}
\definecolor{edits}{HTML}{000000}

\begin{document}

\definecolor{activesearch}{HTML}{D35C01}
\definecolor{investigated}{HTML}{17B890}
\maketitle
\begin{abstract}
The increasing integration of artificial intelligence (\ai) in visual analytics (\va) tools raises vital questions about the behavior of users, their trust, and the potential of induced biases when provided with guidance during data exploration. We present an experiment where participants engaged in a visual data exploration task while receiving intelligent suggestions supplemented with four different transparency levels. We also modulated the difficulty of the task (easy or hard) to simulate a more tedious scenario for the analyst. Our results indicate that participants were more inclined to accept suggestions when completing a more difficult task despite the \ai's lower suggestion accuracy. Moreover, the levels of transparency tested in this study did not significantly affect suggestion usage or subjective trust ratings of the participants. Additionally, we observed that participants who utilized suggestions throughout the task explored a greater quantity and diversity of data points. We discuss these findings and the implications of this research for improving the design and effectiveness of \ai-guided \va\ tools.

\ccsdesc[500]{Human-centered computing~Visual analytics}
\ccsdesc[300]{Human-centered computing~Empirical studies in visualization}

\printccsdesc   
\end{abstract} 

\newif\ifnotes
\notestrue
\newcommand{\acro}[1]{\textsc{\MakeLowercase{#1}}}
\newcommand{\jen}[1]{\ifnotes{\leavevmode\color{purple}{[JH: #1]}}\else{#1}\fi}
\newcommand{\ao}[1]{\ifnotes{\leavevmode\color{teal}{[AO: #1]}}\else{#1}\fi}
\newcommand{\sm}[1]{\ifnotes{\leavevmode\color{orange}{[SM: #1]}}\else{#1}\fi}
\definecolor{control}{HTML}{94C1DC}
\newcommand{\control}[0]{\textbf{\textcolor{control}{control}}}

\newcommand{\doticon}[0]{\includegraphics[width=.014 \textwidth]{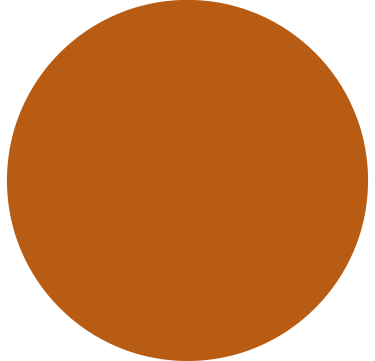}}
\newcommand{\conficon}[0]{\includegraphics[width=.06 \textwidth]{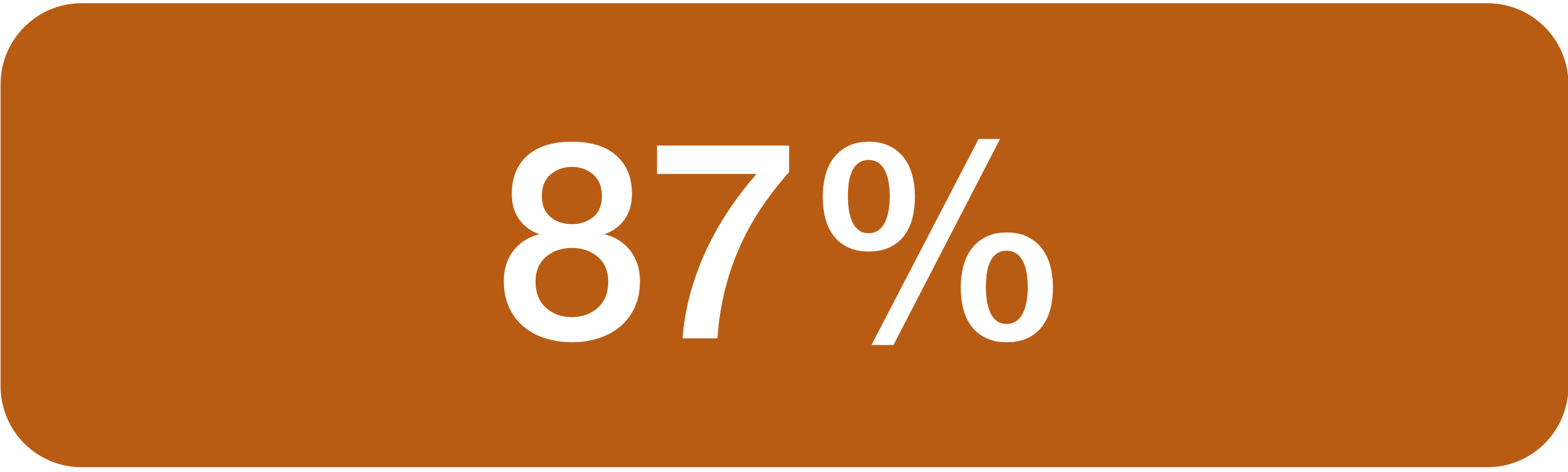}}
\newcommand{\keyicon}[0]{\includegraphics[width=.06 \textwidth]{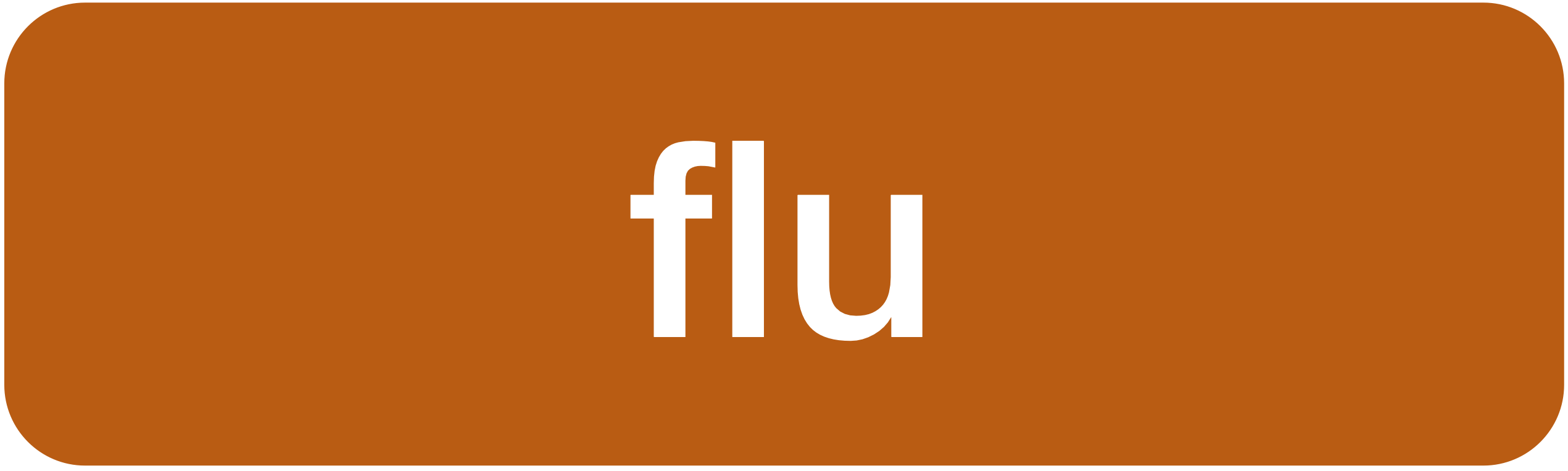}}
\newcommand{\confkeyicon}[0]{\includegraphics[width=.065 \textwidth]{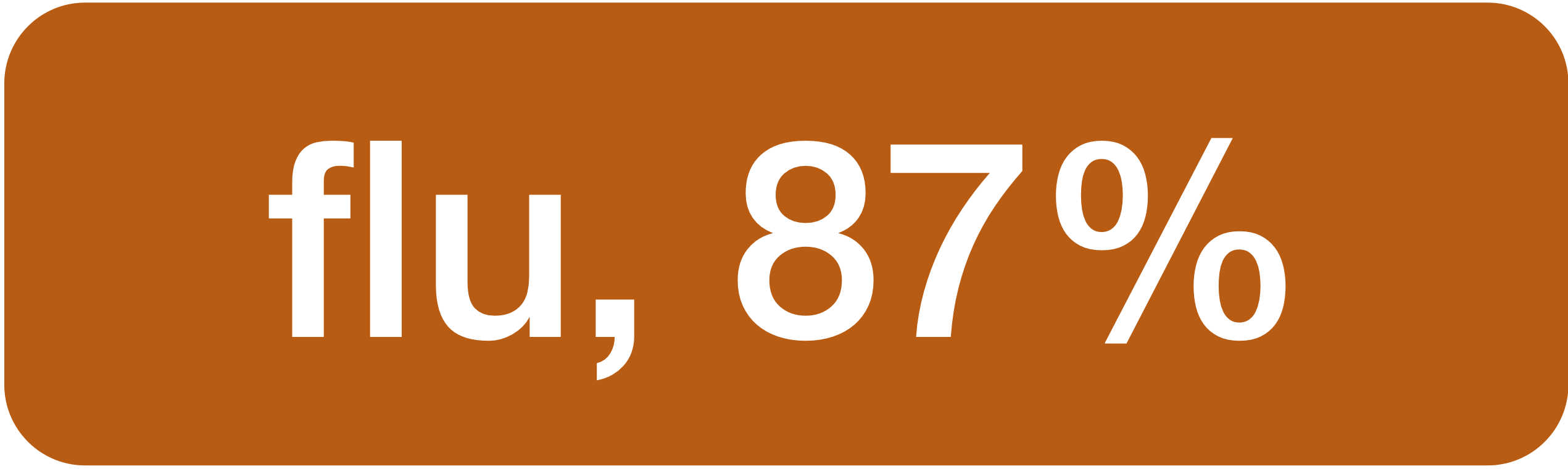}}

\section{Introduction}
The visual analytics (\va) community is increasingly interested in using artificial intelligence (\ai) algorithms to assist users in data exploration and decision-making. As these mixed-initiative systems become more prevalent, it is critical to understand the users' behaviors when interacting with the algorithm and to what extent they allow the algorithm to influence their data exploration and analytical decisions. Understanding the underlying factors that influence the user's interactions is essential to ensure that the \ai\ guidance is used effectively and will maximize the cost-benefit ratio of creating such tools. However, recent studies show two extreme behaviors of interaction within human-\ai\ decision-making. Some users are unable or unwilling to trust an algorithm’s guidance, leading to \textit{underutilization}~\cite{monadjemi2022guided}. On the other hand, some may be heavily influenced by the \ai\ and blindly accept incorrect suggestions without verifying whether it was correct, commonly referred to as \textit{overreliance}~\cite{bucinca_trust_2021}.

Developing methods to mitigate these behaviors and allow users to calibrate an appropriate reliance on the \ai\ during decision-making is an ongoing challenge. One option explored by existing work in the \textsc{xai} community is to provide explanations of the \ai\ suggestions to promote trust calibration. These explanations can be as simple as displaying the \ai\ confidence in a certain prediction or more elaborate explanations that relate information such as feature importance~\cite{wang_are_2021}. Researchers posit that these explanatory techniques can create more effective, transparent, and trustworthy systems by explaining how the \ai\ teammate arrived at its suggestions~\cite{ribeiro_why_2016}. As a result, users can better understand the reasoning behind these suggestions and feel more confident using them during their decision-making process~\cite{kunkel_let_2019}. 

Empirical evidence supports these assertions, with prior work showing that providing explanations of the \ai\ suggestions tended to increase people's ability to calibrate trust in the algorithms~\cite{kunkel_let_2019, zhangEffectConfidenceExplanation2020}. Most of the tasks completed by participants in the prior studies utilized a pre-trained \ai\ for decision-making for scenarios such as medical diagnosis~\cite{bussone_role_2015, gaube_as_2021}, recidivism~\cite{zhangEffectConfidenceExplanation2020}, and income prediction~\cite{ma_who_2023}. Observing the effects of explanations on the user's trust, bias, and behaviors is underexplored in the context of \ai-guided visual data exploration. The exploratory nature of \ai-guided \va\, along with the already utilized visual elements to convey data, adds another layer of complexity to this open challenge. Displaying too much information can overwhelm users and lead to confusion, especially in data-rich settings, which calls the \va\ community for design guidelines on effectively showing and explaining suggestions whilst exploring.

Motivated by these challenges, we explore the impact of task difficulty and transparency on users' trust and data exploration with \ai\ suggestions within a \va\ scenario. 
We conducted a 2$\times$4 between-subject crowdsourced user study with 500 participants tasked with exploring a dataset using a mixed-initiative system. We randomly assigned participants to one of ten groups selected from two task difficulty conditions (easy and hard), four transparency conditions (no transparency, confidence, keyword, keyword + confidence), and two control conditions (no \ai) for each task difficulty group. 

To mimic an ecologically valid analytic task in the national security domain, participants explored a map visualization containing geo-tagged social media posts to identify people who may be affected by an epidemic. They interacted with the dataset for up to ten minutes by tagging individuals of interest, while also being mindful of the different types of symptoms and affected regions of the city. The \va\ system for our study utilized an active search algorithm introduced by Monadjemi et al.~\cite{monadjemi2022guided}, which learns the most influential keyword of interest based on the user’s data exploration interaction logs and recommends data points that will potentially match their analytic goals. 

We find that when completing a more difficult task, the users were more inclined to accept the \ai\ suggestions despite the \ai\ having lower suggestion accuracy. Also, we demonstrate that the level of transparency had no measurable impact on suggestion usage and trust. Moreover, the participants who received no transparency exhibited similar performance measures and subjective trust levels as those who were provided with model transparency. Overall, we found that participants tended to trust \ai\ suggestions regardless of the transparency level. Furthermore, participants overrelied on the suggestions in a more difficult task setting. Lastly, we observed that the participants who utilized the suggestions explored a greater quantity and diversity of data points. We conclude by discussing the implications of our results, highlighting some of the promises and challenges of promoting transparency in difficult task settings along with \ai\ suggestions in a guided visual data discovery framework.

\noindent A summary of our contributions is as follows:
\vspace{-0.5em}
\begin{itemize}[topsep=0pt]
    \item Despite long-standing beliefs about trust and transparency in the \ai\ community, we show that additional transparency to a \va\ system may not always affect data exploration and suggestion usage. We discuss future directions for transparency methods and the unique consideration for \va\ systems.
    \item We demonstrate that users' baseline trust in \ai-guided \va\ may be high, and adding information to promote transparency may have a marginal impact. Still, our findings provide weak confirmation for suggestion usage as a proxy for real-time trust in future \ai-assisted \va\ evaluations as we observe that higher \ai\ reliance was associated with high perceived trust.
    \item We show that task difficulty was strongly associated with suggestion usage and overreliance. Our findings suggest that designers of \va\ systems should consider ways to adapt the level of guidance provided by the \ai\ based on the task's difficulty.
\end{itemize}

\section{Related Work}
Guidance in \va\ is a computer-assisted process to address users' knowledge gaps during interactive sessions~\cite{ceneda_characterizing_2017}. Researchers have developed various systems that utilize \ai\ and \ml\ algorithms ~\cite{xu2020survey} to support user interactions and the discovery of new insights during data exploration~\cite{kery_towards_2019,lin_rclens_2018,monadjemi2022guided}. We contribute to the body of work on guidance in \va\ by examining the relationship between trust, suggestion usage, data exploration, and transparency levels.

\subsection{Evaluating Guidance in \va}
Evaluating guidance in \va\ systems typically involves calculating and comparing metrics such as task accuracy and speed to determine the effectiveness of assisting the user~\cite{monadjemi2022guided, brown_dis-function_2012, battle_dynamic_2016}. While these metrics are crucial, observing how users interact with the \ai\ guidance is equally important to improve the design of future \ai-guided \va\ systems. 

Some existing work has sought to look beyond speed and accuracy \cite{dabek_grammar-based_2017,lee_deconstructing_2021,dasgupta_familiarity_2017,monadjemi2022guided}. For example, work by Dabek et al.~\cite{dabek_grammar-based_2017} examined \textit{suggestion usage}. They found that users ultimately performed the action suggested to them 20\% of the time in their system's evaluation. Their users said the suggestions were useful but not always necessary to solve the task. Lee et al.~\cite{lee_deconstructing_2021} also investigated usage with \textit{Frontier}, a system recommending new ways to visualize a given dataset. They observed that users followed suggestions while exploring unfamiliar data attributes or when they did not know what to explore next. Furthermore, they argue that the interpretability of the suggestions positively impacted usage. Dasgupta et al.~\cite{dasgupta_familiarity_2017} examined \textit{trust perception}. They presented a comparative study of domain scientists' trust level in their visual analytics system, Active Data Biology. They argue that domain scientists trust intuitive and transparent systems that allow seamless switching between hypothesis generation and evidence gathering.


\vspace{-0.8em}
\subsection{\ai\ Underutilization and Overreliance}

Skepticism and low self-reported trust in \ai\ have led users to ignore the \ai\ during the decision-making process and develop algorithmic aversion~\cite{monadjemi2022guided, dietvorst_algorithm_2015, kim_when_2023}, which is defined as the tendency for users to discount suggestions from \ai\ more heavily than human suggestions. This negative attitude towards \ai\ leads to \textit{underutilization} of intelligent suggestions. Researchers have also observed this behavior when users had high-domain expertise~\cite{dabek_grammar-based_2017}. Closely related to domain expertise is task familiarity. Interestingly, another study showed that users with high task familiarity reported more trust in the \ai\ teammate but showed less adherence to its suggestions~\cite{schaffer_i_2019}. 

On the opposite extreme of underutilization is \textit{overreliance}. The literature on \ai-assisted decision-making has also established that humans can be easily influenced by the \ai\ teammate and often accept incorrect suggestions without verifying whether the \ai\ was actually correct~\cite{bucinca_trust_2021}. Jacobs et al.~\cite{jacobs_how_2021} found that users with low \ai\ literacy were significantly more likely to select medical treatments that were aligned with the \ai\ suggestions. These interaction behaviors depend on a lot of different factors, however, the most popular method used to mitigate these behaviors in existing works is to provide explanations of the \ai\ suggestions~\cite{vasconcelos2022explanations, zhangEffectConfidenceExplanation2020, wang_are_2021, nourani_role_2020}. The intuition runs that if users see an incorrect explanation, they will more carefully scrutinize the \ai\ suggestion and build an appropriate reliance.

\subsection{\ai\ Transparency and Explanations}
\label{sec:explanations}

Explainable \ai\ (\textsc{xai}) are techniques that enable humans to understand, trust, and manage \ai\ teammates effectively~\cite{arrieta2020explainable}. 
For example, Cheng et al.~\cite{chengExplainingDecisionMakingAlgorithms2019} 
show that presenting the inner workings of a university admissions algorithm with an interactive interface can enhance users' understanding of the algorithm. However, it is unclear what criteria constitute a sufficient explanation. Researchers have used \textit{examples}~\cite{caiEffectsExamplebasedExplanations2019, yang_how_2020} and \textit{counterfactual examples}~\cite{wang_are_2021}. Wang et al.~\cite{wang_are_2021} explored other explanation methods such as \textit{feature importance}, \textit{feature contribution}, and \textit{nearest neighbors} for recidivism prediction and forest cover prediction tasks. Their study found supportive evidence suggesting that providing information about feature contribution allowed participants to have an awareness of uncertainty within the model and appropriately calibrate their trust.

Transparency and explanations are often interconnected. A common approach to promote transparency is showing \textit{uncertainties}~\cite{helldinPresentingSystemUncertainty2013, zhangEffectConfidenceExplanation2020, bussone_role_2015} of the \ai. For instance, Zhang et al.~\cite{zhangEffectConfidenceExplanation2020} studied the effect of showing \textit{confidence score} and local explanation for predictions in an income prediction task. Their findings show that prediction-specific confidence information could support trust calibration. Similarly, Dietvorst et al.~\cite{dietvorst_overcoming_2018} communicated a \textit{model’s uncertainty} through the outright disclosure of the model’s average error rate by a text description (i.e., “the model has an average error rate of x”). 
Linder et al.~\cite{linderHowLevelExplanation2021} explored how the type and amount of explanations affect users’ understanding and performance on a fact-checking task. More detailed explanations such as providing examples of alternative statements with the same classification and information about the influence of the statement's metadata led the users to a better understanding of the \ai\ suggestions. However, this was coupled with lower performance due to the additional time and attention required.  

\vspace{-1.66em}
\subsection{Manipulating Model Uncertainty and Task Difficulty}

High levels of \ai\ prediction uncertainty may indicate an elevated likelihood of poor performance. Sacha et al.~\cite{sacha_role_2016} 
discussed the role of uncertainty, awareness, and trust in \va. They argue that the users' trust in the \ai\ teammate's outcomes is influenced by their awareness of the various kinds of uncertainty that exist or are generated in the system. To this end, prior works have manipulated task difficulty or model uncertainty whilst providing explanations to observe their respective effects on the human's interactions with the \ai. However, the results are inconclusive. 

For example, Vasconcelos et al.~\cite{vasconcelos2022explanations} manipulated the task's difficulty with a visual search maze-solving task by changing the maze's dimensions such that the harder task involved solving a higher-dimensional maze. They show that explanations of the \ai\ prediction reduced overreliance in the hard task condition. On the other hand, Zhao et al.~\cite{zhao_evaluating_2023} explored the impact of uncertainty visualization on trust and reliance on model predictions using a college admissions forecasting task. 
Findings show that in low-uncertainty tasks, proper visualization of model uncertainty can enhance an appropriate adoption of model predictions. However, when a decision task had high model uncertainty, the uncertainty visualization did not significantly affect the participants' trust.

The overarching findings from some of the prior work support that local explanations and transparency methods may increase trust and suggestion usage. However, these techniques may not necessarily increase human-\ai\ performance. For example, too much information about the \ai\ could lead to worse performance or decision-making outcomes~\cite{linderHowLevelExplanation2021}. 



\section{Research Goals}\label{sec:goals}
Building on prior work, this paper aims to understand how different levels of transparency provided by \va\ systems impact users' trust and acceptance of those suggestions. We are also interested in how the decision to utilize or not utilize these suggestions affects users' data exploration patterns. Further, we consider how task difficulty, \textit{determined here by the percentage of irrelevant data in the underlying dataset}, might further affect the observed behaviors. 

The interplay among trust, suggestion usage, transparency levels, and task difficulty carries significant implications for future \va\ tools seeking to harness \ai\ technology to assist users in data exploration, making the work in this paper a compelling unresolved research inquiry. It is important to consider the prior work showing that the quality of the explanations provided by the \ai\ plays a pivotal role~\cite{lai_human_2019, wang_are_2021} in the users' interactions with \ai\ suggestions. When the \ai\ provides explanations or displays information about its confidence, users are likelier to trust and rely upon \ai\ suggestions~\cite{zhangEffectConfidenceExplanation2020}. Moreover, the intricate relationship between trust and suggestion usage appears to be influenced by task difficulty or model uncertainty~\cite{vasconcelos2022explanations, zhao_evaluating_2023}. We hypothesize that users may be more inclined to trust and utilize \ai\ suggestions when confronted with difficult tasks where additional guidance in data exploration becomes more essential. 




\begin{figure*}[!ht]
    \centering
    \includegraphics[width=0.7\linewidth]{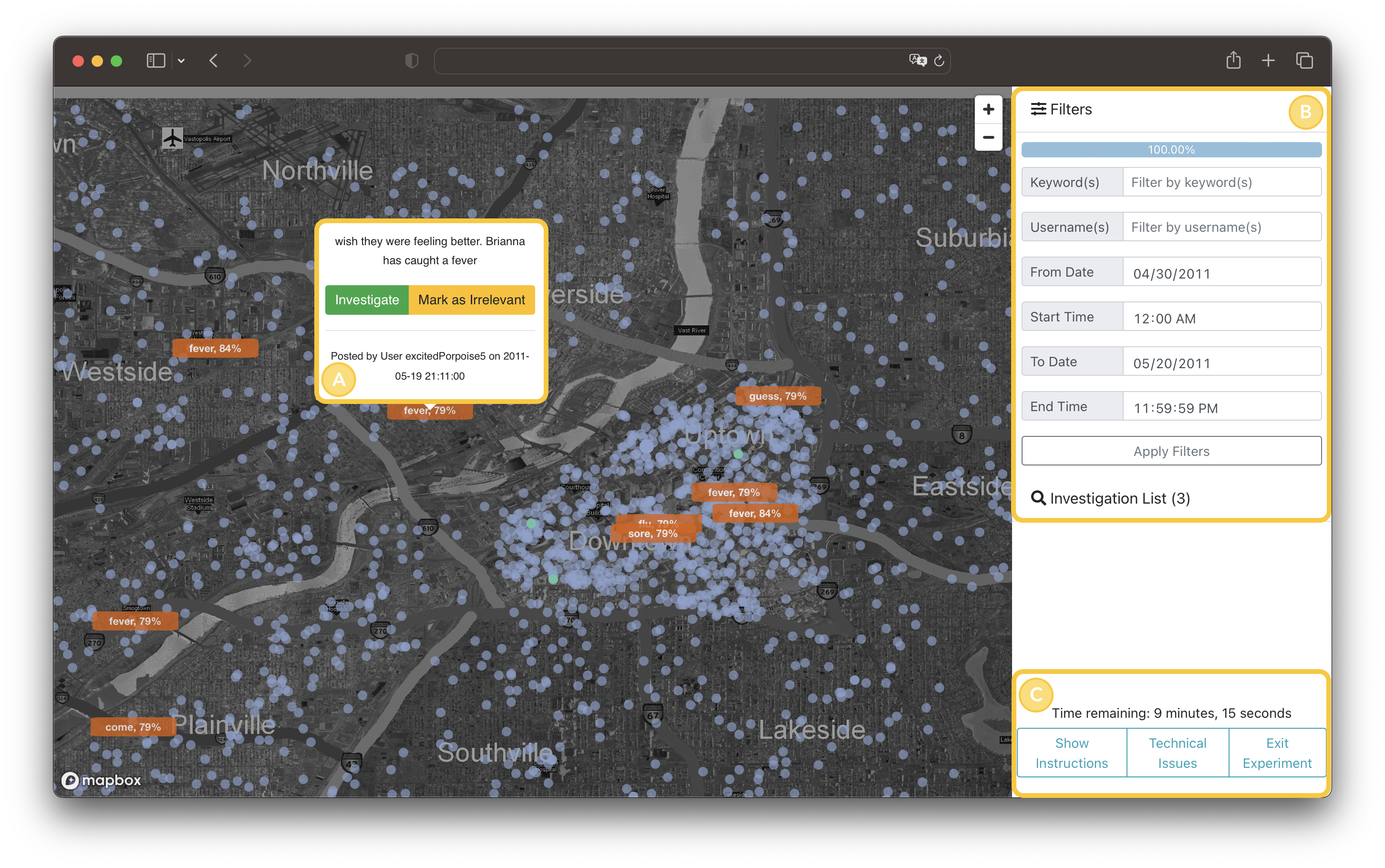}
    \caption{The interface of the system used for the study. a) Hovering triggers a tooltip with the full text of the social media post. b) The sidebar allows users to search and filter. c) The remaining time for the task, and users could exit at any time or report technical issues.}
    \label{fig:vastapolis_map}
    \vspace{-1em}
\end{figure*}

\section{Methods}\label{sec:experiment}
We aimed to create a realistic dataset, scenario, and tasks for our participants. Thus, we adopted the task and dataset from the 2011 Visual Analytics Science and Technology (\textsc{vast}) Challenge. The \textsc{vast} Challenge is a yearly competition and workshop supported by \textsc{ieee vis} and the Pacific Northwest National Laboratory and publishes datasets and analytic tasks that mimic real-world challenges\cite{cook2012vast,cook2014vast}. In this scenario, the fictional city of Vastopolis is facing a biochemical attack, leading to an epidemic that is spreading throughout the city. The dataset consists of 1,023,077 messages, similar to tweets, that were posted on social media from different parts of the city over 21 days. Additionally, there is a satellite image of the city that includes labeled highways, hospitals, landmarks, and water bodies. 

\subsection{Task}
\label{ss:task}
As part of the study, we informed the participants that hospitals in the city of Vastopolis had a significant rise in reported illnesses. The city authorities have recruited the participants to help identify the affected areas by analyzing social media activity. Their task was to sift through the dataset of social media posts using the interface in Figure~\ref{fig:vastapolis_map} and tag posts containing illness-related content for further investigation. Based on their analysis, they then identified the areas of the city that they believed were most impacted by the epidemic. We chose this dataset and task for two main reasons. First, it mimics realistic scenarios where users search through a large space in search for relevant data points (e.g., material discovery, intelligence analysis)~\cite{jiang2018efficient,garnett2012bayesian}. Secondly, the task does not require any domain expertise, which makes it appropriate for a large-scale crowd-sourced study targeted towards the general public.


\subsection{Design}
We designed a 2 (task difficulty) $\times$ 4 (transparency level) between-subject study, resulting in a total of eight \ai\ conditions. There were two control (no \ai) conditions for each level of difficulty.  

\subsubsection{Task Difficulty} 
Controlling the difficulty of the task aids in understanding when interactions with the \ai\ change and when guidance may be valuable but more error-prone. To control the level of difficulty of our task, we use the percentage of irrelevant points in the dataset as a proxy. The study had two levels of difficulty:
\vspace{-0.5em}
\begin{itemize}[topsep=0pt]
    \item \textbf{\hard}: The data was a random sample of 2000 points from the entire 21-day period, resulting in $\sim$9\% illness-related posts.
    \item \textbf{\easy}: The data was a random sample of 2000 points from the approximate start date of the epidemic. Hence, adopting the assumption that the starting point of the epidemic is known. This resulted in a dataset with $\sim$36\% illness-related posts.   
\end{itemize}


\subsubsection{Transparency Levels}
\label{sec:explanation}


With limited screen real estate in \va\ and the added stress of solving a task within a limited time, we need to be conservative in how we promote situational awareness of the \ai\ and how much information is shown to the users, especially when the \ai\ provides multiple suggestions. Although the space of explanation techniques is vast, we opted for variations of simple prediction-specific transparency methods for users to understand -- the most influential keyword and confidence value. Inspired by prior studies~\cite{zhangEffectConfidenceExplanation2020, linderHowLevelExplanation2021, wang_are_2021}, we posit that providing the most influential keyword and confidence value may improve trust calibration by giving users an indication to increase their situational awareness of the \ai's performance. We define an influential keyword to be one which if eliminated, decreases the probability of a post being relevant by the largest amount. This method of transparency will allow users to quickly understand the context of a certain suggestion, while confidence values can provide users with information about the system's level of certainty regarding the relevance of a particular suggestion. In addition to a \textbf{control (\ctrl\ | no \ai)} condition, there were four transparency levels in our study:
\vspace{-0.5em}
\begin{itemize}[topsep=0pt]
    \item \textbf{No transparency (\none)}: Suggestions are presented as dots. \doticon
    \item \textbf{Confidence (\conf)}: Suggestions are presented as rectangles containing the \ai's confidence percentage that the suggestion is relevant to the user. \conficon
    \item \textbf{Keyword (\key)}: Suggestions are presented as rectangles containing the most influential word for classifying the social media post as relevant to the user. \keyicon
    \item \textbf{Keyword + confidence (\keyconf)}: Suggestions are presented as rectangles containing the most influential word and the \ai's confidence percentage. \confkeyicon
\end{itemize}

\subsection{Visual Analytic System}
We expanded upon the prototype presented in~\cite{monadjemi2022guided} and prioritized the following design features for the \va\ tool used in this study: (1) a map for identifying regions that are most affected, (2) a search and filter functionality to aid the discovery of illness-related terms (3) the ability to inspect individual social media posts and triage people for contact tracing. See~\autoref{fig:vastapolis_map} for the tool's interface. 
Users hovered over data points to trigger a tooltip with the social media post's details. They then could flag the post by clicking the \textit{investigate button}. If the post was previously flagged, there is an option to remove the flag and report an \textit{irrelevant suggestion} from the \ai. We utilized three distinct colorblind-safe colors to distinguish among \textbf{\textcolor{activesearch}{suggested points}}, \textbf{\textcolor{investigated}{points selected for investigation}}, and the \textbf{\textcolor{control}{remaining points}}.


\subsection{Modeling and Guidance Engine}\label{ss:model}
Our visual analytic system observed user interactions, inferred their data interest, and made suggestions. To fully specify our guidance engine, we need to describe (1) the classification model which \emph{predicted} the relevance of documents to the task at hand in light of past interactions, and (2) the algorithm which \emph{decided} which documents to suggest given the classification model's belief. 

\subsubsection{Predicting Document Relevance}

We begin with a finite set of $n$ data points displayed on the interface, \mbox{$\mathcal{X} = \{x_1, x_2, ..., x_n\}$}, and as we observe user interactions, we wish to infer if each point in $\mathcal{X}$ is relevant to the task at hand. We consider this dataset to be \emph{unlabeled} initially, meaning that we do not know whether each point is relevant to the task at hand or not. As we observe user interactions with data points, we translate their interactions into labels and collect them in the observation set, \mbox{$\mathcal{D} = \{(x_1, \hat{y}_1), ..., (x_t, \hat{y}_t)\}$}, where $\hat{y}_i=1$ indicates that $x_i$ is deemed \emph{relevant} and $\hat{y}_i=0$ indicates that $x_i$ is deemed \emph{irrelevant} based on user interactions. We want to then reason about the relevance of the remaining points to the task at hand, \mbox{$\Pr(\hat{y}_i=1 \mid x_i, \mathcal{D})$}, where $x_i \in \mathcal{X}$ is an arbitrary unlabeled data point.
Specific to our visual analytic tool described above, we consider an \emph{investigate} interaction to result in a $\hat{y}_i=1$ label (i.e., relevant), whereas \emph{irrelevant suggestion} interactions result in a $\hat{y}_i=0$ label (i.e., irrelevant).

Since our task and dataset from Section \ref{ss:task} involve interacting with social media posts, we wish to express $\Pr(\hat{y}_i \mid x_i, \mathcal{X})$ with support over text data. Therefore, in a pre-processing step, we transform the unstructured text data into a numerical space using a pre-trained $word2vec$ auto-encoder \cite{rehurek_lrec}. After this transformation, each social media post is represented as a 300-dimensional numerical vector. We then compute the pairwise cosine distance between vectors to build a $k$-\acro{NN} classification model for inferring the relevance of documents to the task at hand in light of past observations where $k = 180$. This model is initially untrained, meaning it does not have any labeled observations. As the user interacts with the data, we re-train this model with the updated observations. One compelling benefit of using a $k$-\acro{NN} classifier in this setting is that it is quick to re-train in real-time. 

\begin{table}[!h]
    \centering
    \caption{Summary of notations used in Section~\ref{ss:model}}
    \begin{tabularx}{\linewidth}{lX}
    \toprule
    Notation & Description \\
    \midrule
     $\mathcal{X}$ & Set of $n$ data points displayed on the interface. \\
     $\mathcal{D}$ & Set of data points and their relevance labels observed through user interactions with the interface. \\
     $\mathcal{S}_t$ & Set of data points suggested to the user at time $t$. \\
     \midrule
     $\hat{y}_i$ & Label for data point $x_i$ observed from interactions. \\
     $y_i$ & Ground-truth label for data point $x_i$ given the task. \\
    \bottomrule
    \end{tabularx}
    \label{tab:notations}
    \vspace{-1em}
\end{table}

\subsubsection{Making Suggestions for Investigation}
We used an active search algorithm to suggest points to the user. Active search is an active learning algorithm that makes queries to maximize the number of relevant data points discovered under a limited querying budget \cite{pmlr-v70-jiang17d}. In mathematical notation, the utility of $\mathcal{D}$ at the end of the session is defined as:
\[
u(\mathcal{D}) = \sum_{j=1}^{\mid\mathcal{D}\mid} \hat{y}_j,
\]
where the active search algorithm aims to make queries to approximately maximize this utility. 
Active search has proven effective in accelerating drug discovery \cite{jiang2018efficient} and visual data foraging\cite{monadjemi2022guided}. We utilized a greedy active search algorithm, assuming each set of queries is the final batch. 
We refer the readers to Jiang et al.~\cite{jiang2018efficient} for details on search horizon and exploration/exploitation trade-off in active search algorithms.

\begin{table*}[ht!]
    \centering
    \caption{Summary statistics and self-reported demographics of participants.}
    \begin{tabular}{lcc}
        \toprule
          & Easy Task & Hard Task\\
         \midrule
         Recruited participants & 236 & 236 \\
         Excluded participants & 44 & 36 \\
         Included participants & 192 & 200 \\
         Sex & 91 Male, 95 Female, 6 Undisclosed & 94 Male, 101 Female, 5 Undisclosed  \\
         Age & $\mu = 34.7$, $\sigma = 12$ & $\mu = 32.7$, $\sigma = 11.1$  \\
         Education & 68\% with at least an associate degree  & 60\% with at least an associate degree\\
         \bottomrule
    \end{tabular}
    \label{tab:demographics}
    \vspace{-1em}
\end{table*}

\subsection{Participants}
\textcolor{edits}{We used G*Power to conduct an a priori power analysis to estimate the sample size for this study. Our effect size of .6 for an \textsc{ANOVA} was based on studies by Monadjemi et al.~\cite{monadjemi2022guided}.
The result from the a priori analysis indicated that we needed a sample size of $N=450$ to achieve 80\% power for detecting a medium effect at a significance level of $\alpha=.05$. Thus, we used a sample size of $N=500$ to account for exclusions. 
We provide additional information about the study in the preregistration\footnote{This experiment was pre-registered on \href{https://osf.io/efxga}{Open Science Foundation}.}.}

We recruited our participants through Prolific~\cite{palan2018prolific} per Washington University's IRB guidelines. 
Participants were 18 to 65 years old, from the United States, and fluent in English. For more detailed participant demographics, please refer to Table~\ref{tab:demographics}. We provided a base pay rate of \$15.00 per hour and the participants' median time to complete the study was around 15 minutes (including the tutorial, the task, and the survey).


\subsection{Experimental Procedure}
Our system randomly assigned each participant to one of the eight conditions. Upon giving consent to participate in our study, participants saw a tutorial demonstrating how to interact with the system. 
\textcolor{edits}{The tutorial also provided explanations as well as examples of the transparency methods tested in this study to all participants.} 

Each participant then played the role of a triage investigator and needed to find social media posts that contained information about the hot spot locations of the epidemic and the symptoms being reported. Participants could either use the search feature or browse the data points via the interactive map shown in Figure~\ref{fig:vastapolis_map}. Hovering over a data point triggered a tooltip containing the post's full text, and clicking on the \textit{investigate} button within the tooltip tagged the post as containing illness-related information. Once the \ai\ observed the participant's first three tagged interactions, it suggested 10 relevant posts to explore. These suggestions were in the form of visual cues that were updated after every investigation. 
The visual cues depended on the type of transparency condition they were assigned as detailed in Section~\ref{sec:explanation}. 

After spending up to 10 minutes tagging posts related to the epidemic, participants answered some of the \textsc{vast} Challenge questions, which included their beliefs on whether the epidemic was contained, how it was transmitted, and the areas that were most affected. Finally, they completed the exit survey which included demographic and system usability questions, in addition to attention checks and questions about their level of trust towards the \ai. 


\section{Data Collection, Exclusions, and Validation}\label{sec:datacleaning}
There were two exclusion criteria listed on our preregistration for this study. First, we initially excluded participants who had technical issues and could not complete the task ($n=28$), then we excluded the sessions of those who failed to pass the attention checks in the post-experiment survey or had less than 10 hovers on the data points in their interaction log ($n=64$). In the end, we had a total of 408 participants' sessions for data analysis.

\subsection{Manual Coding and Ground Truth Proxy}

In addition to measuring trust and suggestion usage, we considered accuracy to enable comparisons with prior work.
However, the 2011 \textsc{vast} Challenge dataset did not include labels indicating posts that were illness-related. We began by compiling a list of all the tagged social media posts from the study participants, producing 1642 unique posts. One author manually coded each post using binary labels of \textbf{illness-related} or \textbf{non-illness-related}. This process also produced a collection of word stems, allowing us to programmatically label the full dataset used in the study. We evaluated the labels by iteratively selecting random samples of the dataset, correcting labeling errors, and updating the list of word stems. For example, the following post in the dataset was initially not flagged as illness-related but after the manual coding process, the label was corrected as ``having the sweats" is a potential symptom of having a fever: \textit{``Nicholas has caught a the \textbf{sweats} I hate this."}

These ground truth proxy labels enable us to estimate measures such as the \ai\ and participants' accuracy. This process also uncovered a few participants who did not complete the task as instructed. Specifically, for 16 participants, none of the social media posts they tagged for contact tracing included illness-related information. Thus, we deviated from our preregistration exclusions and excluded these 16 participants from our analysis, resulting in 392 remaining participants. Table~\ref{tab:demographics} summarizes the self-reported demographic information of participants included in the study. 

\subsection{Data Collection}

To measure the impact of task difficulty and transparency level on \ai\ interaction, we calculated the following dependent variables:
\begin{itemize}[topsep=0pt]
    \item \textbf{Suggestion usage} $\in [0...1]$, the ratio of \ai\ suggestion batches that resulted in an interaction (\textit{investigate} or \textit{mark as irrelevant}). \usage\ is the quotient of the number of accepted suggestions divided by the number of \ai\ suggestion batches received. 
    $$\textsc{usage} = \frac{1}{\mid\mathcal{D}\mid-3}\sum_{t=4}^{\mid\mathcal{D}\mid}{\mathbbm{1}_{\mathcal{S}_{t-1}}(x_{t})},$$
    where $\mathbbm{1}$ denotes the indicator functions 
    \footnote{The indicator function states that $\mathbbm{1}_{\mathcal{S}_{t-1}}(x_{t}) = 1$ if $x_{t} \in \mathcal{S}_{t-1}$, and $\mathbbm{1}_{\mathcal{S}_{t-1}}(x_{t}) = 0$ otherwise. In our case, it tells us whether the data point with which the user interacted was suggested to them by the system.}.
    \item \textbf{\ai\ accuracy} $\in [0...1]$, the ratio of relevant suggestions presented to the participant. 
    $$\textsc{ai accuracy} = \frac{1}{\mid\mathcal{D}\mid - 3}\sum_{t=4}^{\mid\mathcal{D}\mid}\sum_{i=1}^{\mid\mathcal{S}_t\mid}{\frac{y_i}{\mid\mathcal{S}_t\mid}}$$
    \item \textbf{No. of positive investigations} is the number of illness-related social media posts tagged for investigation by the participant.
    $$\textsc{investigations} = \mid\{x_i \in \mathcal{D} \mid (y_i = 1) \wedge (\hat{y}_i = 1)  \}\mid$$
    \item \textbf{Overreliance} $\in [0...1]$, the ratio of irrelevant suggestions investigated by the participant.
    $$\textsc{overreliance} = \frac{\mid\{x_t \in \mathcal{D} \mid (x_t \in \mathcal{S}_{t-1}) \wedge (\hat{y}_t = 1) \wedge (y_t=0)\}\mid}{\mid\{x_t \in \mathcal{D} \mid (x_t \in \mathcal{S}_{t-1}) \wedge (\hat{y}_t = 1)\}\mid}$$
    \item \textbf{Symptom diversity} is the number of unique relevant symptoms discovered by the participant through investigations.
\end{itemize}

\noindent
The covariate that was measured:
\begin{itemize}
    \item \textbf{\trust} $\in [1...5]$, self-reported trust in the algorithm collected in the post-experiment survey. Participants responded to the statement ``\textit{I trusted AVA throughout the investigation.}" on a 5-point Likert scale with 1 = Strongly disagree to 5 = Strongly agree.
    
    

\end{itemize}

\section{Hypotheses}
\label{sec:hypotheses}
We tested the following hypotheses:
\begin{itemize}
    \setlength{\labelsep}{1em}
    \setlength{\itemindent}{1.5em}
    \item [\textbf{H1}] Participants in \hard\ will use suggestions more frequently than those in \easy, regardless of the level of transparency. Aligning with prior work which suggests that participants have a tendency to overrely on \ai\ with difficult tasks~\cite{vasconcelos2022explanations}, we hypothesize that participants in \hard\ may be more willing to rely on external guidance for direction throughout their analysis.

    \item [\textbf{H2}] Participants who receive higher transparency (i.e., \keyconf) will have higher suggestion usage and subjective trust ratings~\cite{linderHowLevelExplanation2021, kunkel_let_2019} compared to those who receive lower transparency (i.e., \none, \conf, and \key), regardless of task difficulty. 

    \item [\textbf{H3}] Participants in \hard\ may have a higher level of trust in the system than those in \easy. Building on \textbf{H1}, we hypothesize that participants in \hard\ may perceive the \va\ system as providing valuable guidance in a difficult task. In contrast, participants in \easy\ may perceive the system as unhelpful since the task is doable without additional guidance~\cite{dabek_grammar-based_2017}. 

    \item [\textbf{H4}] Participants in \conf\ and \keyconf\ will have higher trust in the system than those in \none\ and \key, regardless of task difficulty. We posit that confidence values communicate the level of certainty the \ai\ has and the presence of the most influential keyword further clarifies the reasoning behind the suggestions. 
\end{itemize}

\section{Results}
Of the 392 participants, 192 were assigned to \easy, and 200 were assigned to \hard. We begin our analysis by establishing a baseline with our control (\ctrl) conditions, i.e., participants who completed the task with no \ai\ suggestions. 

\noindent
\textbf{Is there evidence that the hard task was more difficult?}\hspace{1em}
Participants in \easy|\ctrl\ $(n = 40)$ had a median value of 70 people tagged for contract tracing (IQR $(40.5, 95.25)$), and those in \hard|\ctrl\ $(n = 45)$ had a median value of 50 (IQR $(27.0, 63.0)$) people tagged. An independent sample Mann-Whitney U test comparing the outputs of the two groups found a significant difference in the number of investigations $(U = 1209, p = .007, \eta^{2} = .087)$. \textit{The data quality in \hard\ made the task more difficult to execute than for those in \easy, eliciting lower discovery rates.} 


\begin{figure}[!h]
    \centering
    \includegraphics[width=0.82\linewidth]{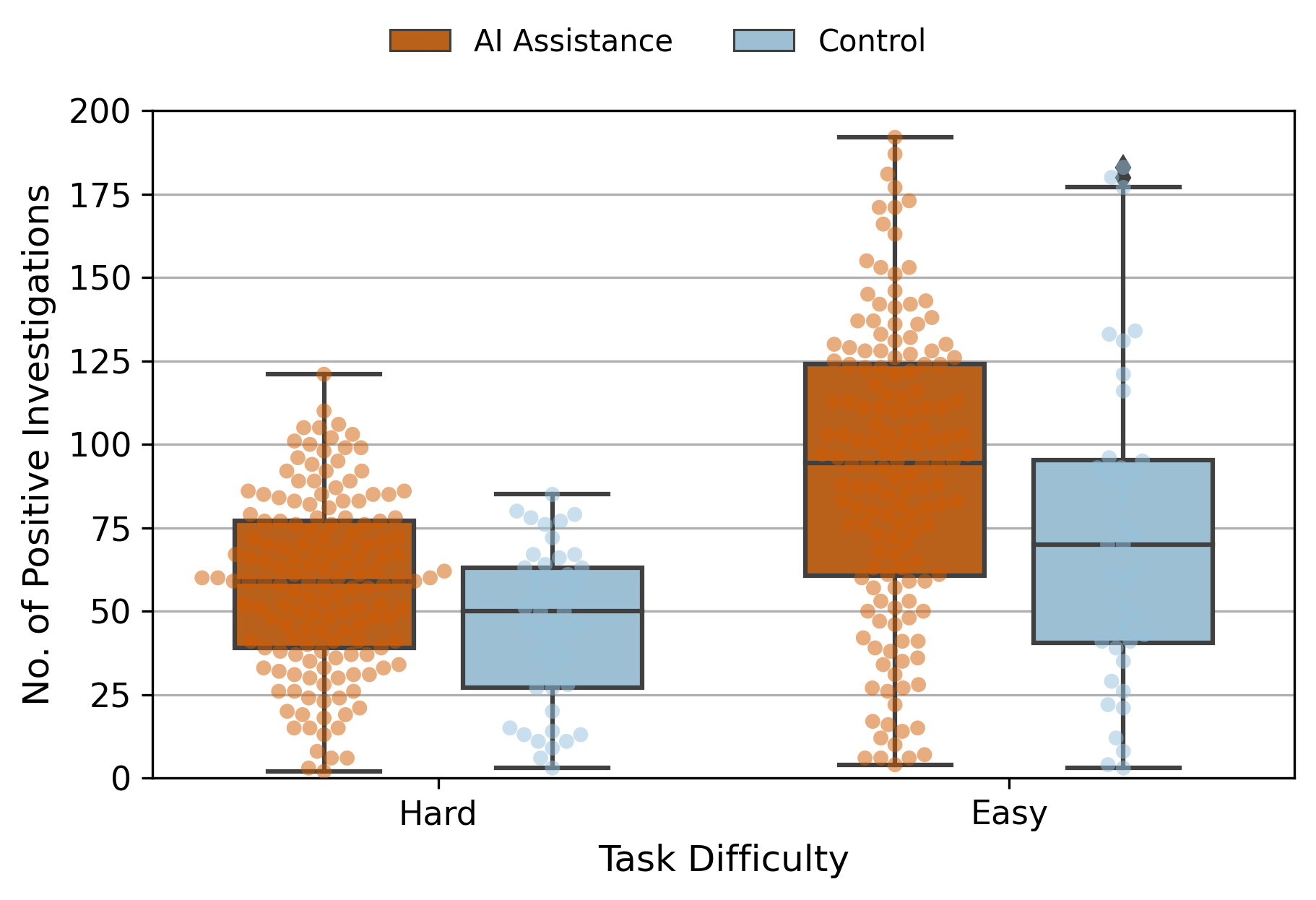}
    \caption{The spread of positive investigations made for \ctrl\ and \ai-guided groups for both \easy\ and \hard. }
    \label{fig:positive_investigations_control}
\end{figure}

\noindent
\textbf{Do we replicate prior work showing that \ai\ improves discovery rate?}\hspace{1em}
Based on prior findings~\cite{monadjemi2022guided}, this work assumes that \ai\ guidance would improve data discovery. 
We validate this assumption by comparing the \inv\ averaged across all \ai\ transparency conditions and the \ctrl (no \ai) condition. Figure~\ref{fig:positive_investigations_control} shows the spread of \inv\ for the \ctrl\ $(n=85)$ and \ai-guided groups $(n=307)$ for both \easy\ and \hard. An independent samples Mann-Whitney U test comparing the outputs of the two groups found a significant difference between the number of people tagged for contact tracing $(U = 16790.5, p < .001, \eta^{2} = .042)$. \textit{This finding replicates prior work~\cite{monadjemi2022guided} and reaffirms that \ai\ guidance is associated with more efficient data exploration.}



\begin{figure}[!h]
    \centering
    \includegraphics[width=0.92\linewidth]{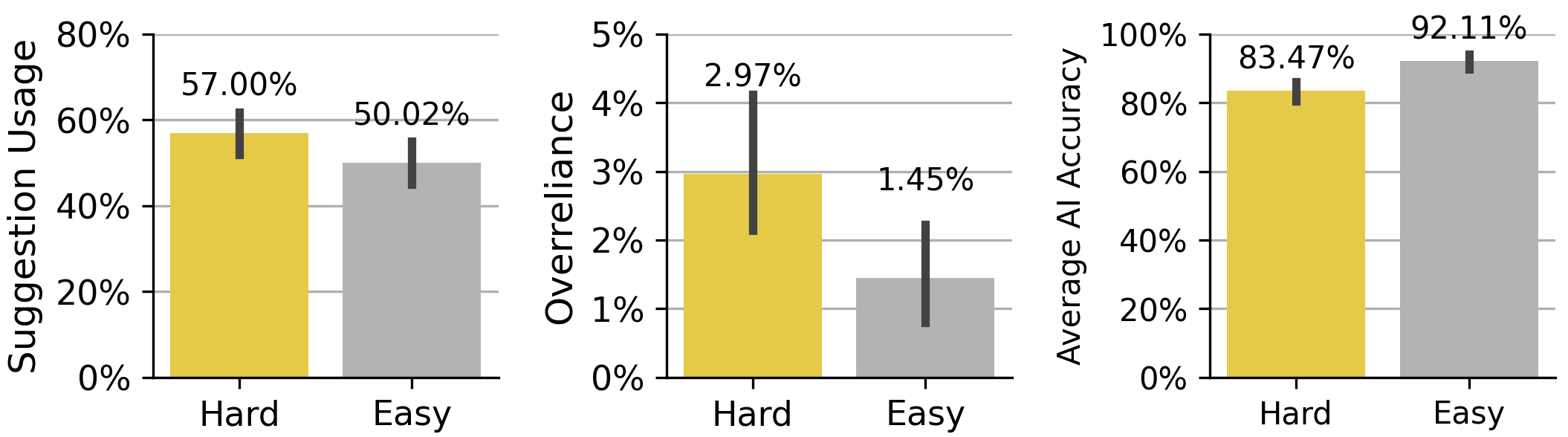}
    \caption{Comparison of suggestion usage, overreliance, and average \ai\ accuracy between task difficulty conditions. Participants in \hard\ utilized more suggestions despite having less accurate \ai\ suggestions, which may have led participants to overrely.}
    \label{fig:sugg_acc_rel}
    \vspace{-1em}
\end{figure} 

\subsection{Is it more likely to use \ai\ suggestions when completing a more difficult task?}
\label{sec:usage_taskdifficulty}



With \textbf{H1}, we hypothesized that participants in \hard\ might feel more uncertain about their ability to perform the task and may be more willing to rely on external guidance. The first chart in Figure~\ref{fig:sugg_acc_rel} compares the average \usage\ between \easy\ (50.02\%) and \hard\ (57\%). 
An independent sample Mann-Whitney U test revealed a significant difference in the \usage\ between the two groups $(U = 10178.5, p = .039, \eta^{2} = .014)$. Supporting \textbf{H1}, we reject the null hypothesis that there is no difference in the proportion of suggestions used between \hard\ and \easy. \textit{Participants in \hard, were more likely to utilize \ai\ suggestions during exploration. However, the effect is small.}

\noindent
\textbf{Is there evidence of overreliance?} \hspace{1em}
The finding that participants in \hard\ were more likely to use \ai\ suggestion than \easy\ is particularly interesting when we consider the \ai\ accuracy. As shown in Figure~\ref{fig:sugg_acc_rel}, we observed a significantly lower \ai\ accuracy for \hard\ compared to \easy\ $(U = 16790.5, p < .001, \eta^{2} = .135)$. Overreliance on \ai\ can lead to blindly accepting incorrect suggestions and potentially harming outcomes. Thus, we examine participants' reliance and whether the difficulty of the task may influence it. Overall, the rate of overreliance was low, with evidence suggesting that, on average, 2.97\% and 1.45\% of participants' accepted \ai\ suggestions in the \hard\ and \easy\ groups were irrelevant to the task. An independent sample Mann-Whitney U test revealed a significant difference in the \textsc{overreliance} distributions between \easy\ and \hard\ $(U = 9315.5, p < 0.0001, \eta^{2} = .033)$. This finding aligns with prior works that observed an \textit{increase in overreliance on the \ai\ when completing more difficult tasks ~\cite{vasconcelos2022explanations}}. The second chart in Figure~\ref{fig:sugg_acc_rel} compares the average overreliance rates between \easy\ and \hard.

\subsection{Is it more probable to use \ai\ suggestions when there is more transparency?} 

\begin{figure}[!h]
    \centering
    \includegraphics[width=0.85\linewidth]{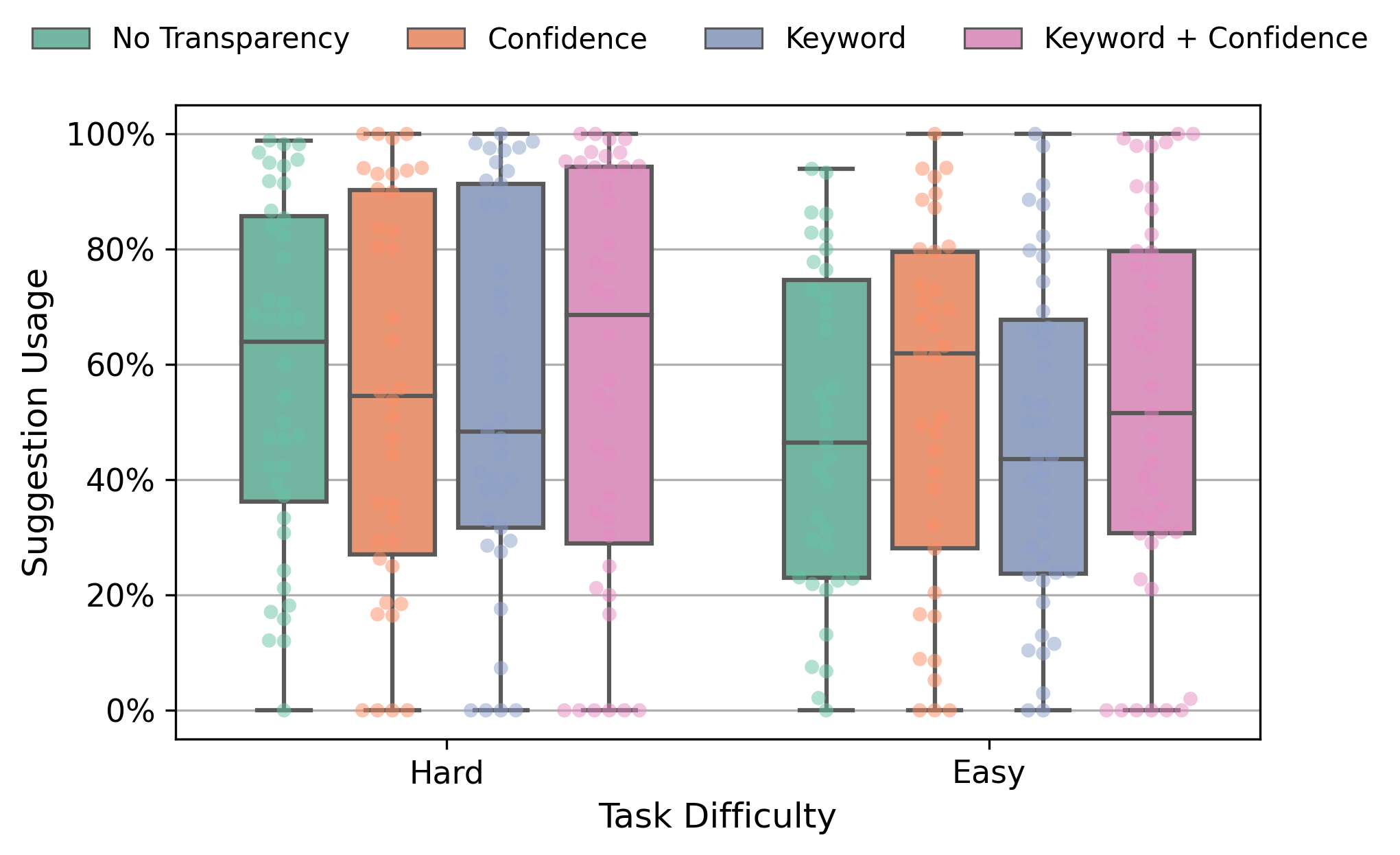}
    \caption{The spread of \usage\ across conditions. We found no significant difference among these groups for both \easy\ and \hard.}
    \label{fig:suggestion_usage}
\end{figure}

Figure~\ref{fig:suggestion_usage} shows the spread of \usage\ among the participants in all eight \ai-guided condition groups.
Despite the median \usage\ for \keyconf\ being higher than all the other conditions within \hard, we fail to reject our null hypothesis that there is no significant difference among the transparency levels for both \easy\ $(H(3) = 1.378, p = .711)$ and \hard\ $(H(3) = .483, p = .923)$. Thus, \textit{contrary to prior work in the \ai\ community~\cite{bansal_does_2021, bucinca_proxy_2020, bucinca_trust_2021, lai_why_2020}, the amount of displayed information did not significantly affect the use of \ai\ suggestions}, opposing \textbf{H2}.

\begin{figure}[!ht]
    \centering
    \includegraphics[width=0.85\linewidth]{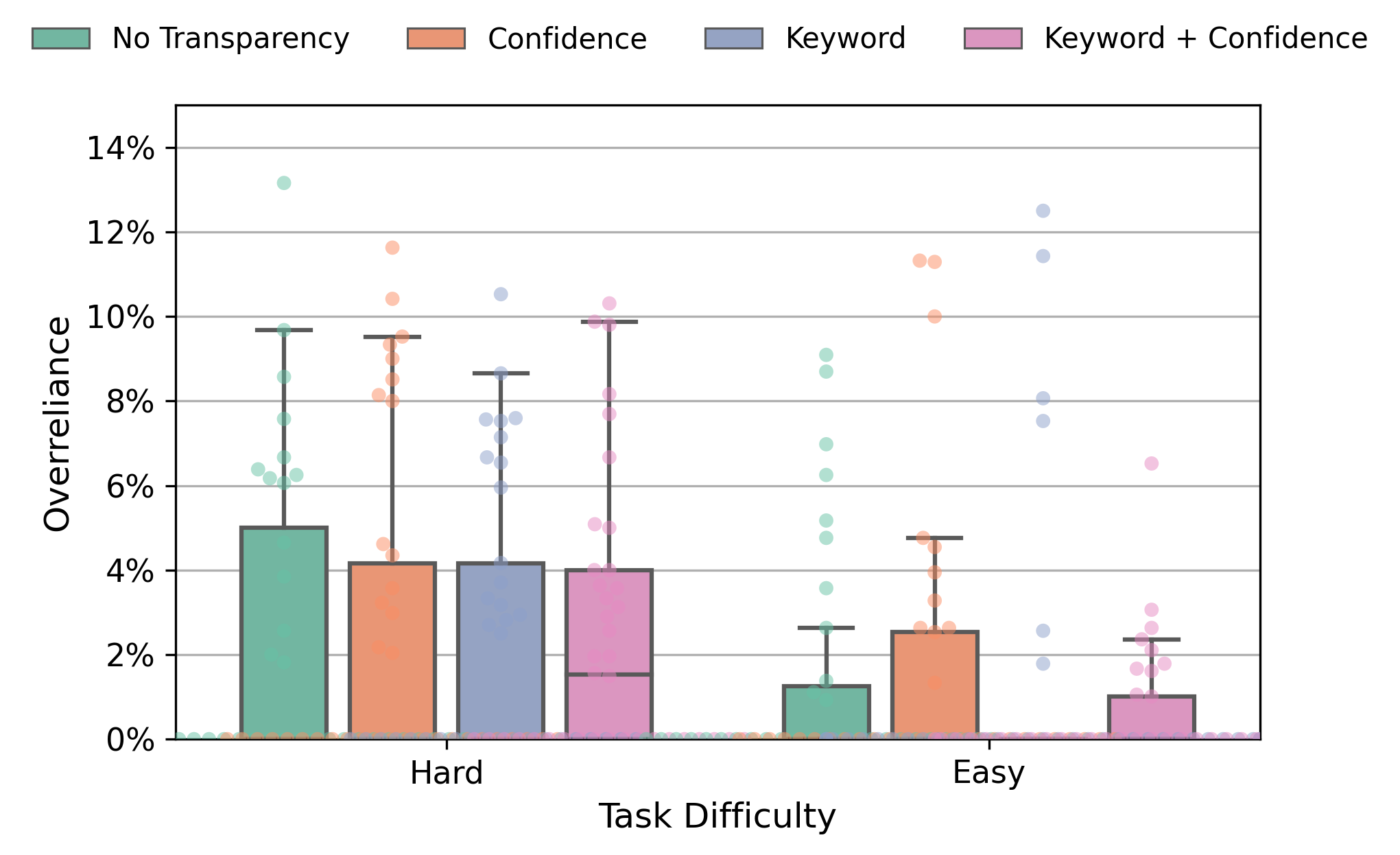}
    \caption{The spread of overreliance across all conditions. Regardless of the transparency condition, participants in \hard\ overrelied on the suggestions at similar rates throughout the experiment.}
    \label{fig:overreliance}
\end{figure}

\noindent
\textbf{Does transparency reduce overreliance?} \mbox{\hspace{1em}}
We ran separate Kruskal-Wallis tests to compare the rate of overreliance for participants within the transparency levels for \easy\ and \hard. Although the spread of overreliance within \conf\ was more dispersed than the rest of the transparency conditions in \easy, we found no significant difference among the transparency levels within both \easy\ $(H(3) = 2.648, p = .449)$ and \hard\ $(H(3) = 913, p = .822)$. Again, contrary to prior work in the \ai\ community, \textit{promoting transparency did not increase the participants' overreliance on the \ai~\cite{bansal_does_2021} nor did it reduce overreliance~\cite{vasconcelos2022explanations}.} Figure~\ref{fig:overreliance} shows the spread of overreliance rates among the participants in all eight condition groups.

\subsection{Does task difficulty affect subjective trust?}

\begin{figure*}[!ht]
    \centering
    \includegraphics[width=0.62\linewidth]{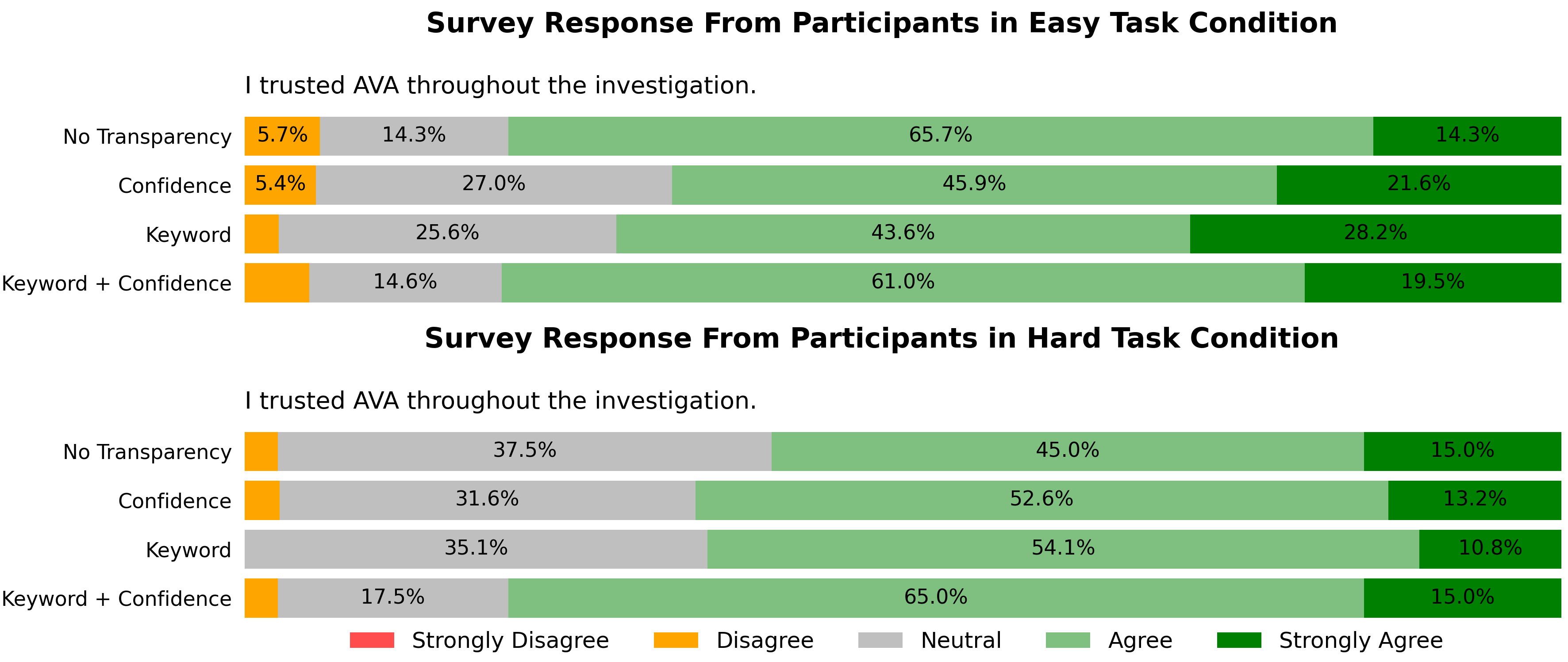}
    \caption{Post-experiment survey responses on trust toward the \ai\ separated by task difficulty and transparency level. We observed high ratings of trust across all conditions with \keyconf\ inducing the highest rating in both \easy\ and \hard.}
    \label{fig:trust}
    \vspace{-1em}
\end{figure*}

Overall, we observed high levels of subjective trust among our participants regardless of the transparency level and task difficulty. Out of all the recruited participants, 71\% of them either agreed or strongly agreed that they trusted the system's suggestions, see Figure~\ref{fig:trust} for a breakdown of self-reported trust among all conditions. With \textbf{H3}, we hypothesized that participants in \hard\ may have a higher level of \trust\ than those in \easy. To test this hypothesis, we compared the distribution of \trust\ between \easy\ and \hard. An independent sample one-tailed Mann-Whitney U test revealed that the distribution of \trust\ levels in the \ai\ for \easy\ was stochastically greater than the distribution of self-reported \trust\ levels of those in \hard\ $(U = 13000.0, p = 0.0417, d = .008)$, which is opposite to what we hypothesized in \textbf{H3}. We find suggestive evidence that \textit{completing a difficult task with \ai\ guidance elicited beliefs that the system was less trustworthy.} 

However, lower \ai\ accuracy in \hard\ conditions, as we saw in Section~\ref{sec:usage_taskdifficulty}, is a confounding factor. We, therefore, considered the correlation between \ai\ accuracy and self-reported trust. A Spearman's rank correlation found no measurable association overall ($r$(305) $= .0491, p < 0.391)$) nor for the \hard\ ($r$(153) $= -.0705, p < 0.383)$).

\noindent
\textbf{Is there a relationship between suggestion usage and trust?}\hspace{1em}
The literature on \ai-assisted decision-making often utilizes suggestion usage as a proxy for trust~\cite{bansal_does_2021,zhangEffectConfidenceExplanation2020, wang_are_2021} \textcolor{edits}{as behavioral trust (measured in this study via suggestion usage) can be defined by an action based on cognitive (or subjective) trust~\cite{kim_how_2021}}. A Spearman's rank correlation was computed to determine whether suggestion usage is a valid proxy for real-time trust in a \va\ system. For both \easy\ ($r$(150) $= .312,  p < .0001$) and \hard\ ($r$(153) $= .245,  p < .002$), there is a weak positive correlation between \usage\ and \trust, \textcolor{edits}{\textit{supporting that suggestion usage has potential to be a valid indicator of subjective trust of users in real-time}}.


\subsection{Does transparency increase subjective trust?}

In \textbf{H4}, we hypothesized that higher transparency would induce higher subjective trust, regardless of task difficulty. We ran separate Kruskal-Wallis tests to compare \trust\ for participants within the transparency levels for both \easy\ and \hard. Our analysis found no significant difference in \trust\ among the transparency levels for both \easy\ $(H(3) = 0.702, p = .873)$ and \hard\ $(H(3) = 2.564, p = .464)$. Thus, we conclude that \textit{promoting \ai\ transparency did not influence subjective trust.} 


\subsection{Does \ai\ guidance induce bias?}
A vital aspect of the \textsc{vast}\ challenge was to generate hypotheses for how and where the epidemic was being transmitted. To answer this question, the participants need to understand the scope of symptoms and identify hotspots of the epidemic. We analyzed how \ai\ guidance impacted the number of relevant symptoms discovered and the distribution of locations explored. 

\begin{figure}[!h]
    \centering
    \includegraphics[width=0.61\linewidth]{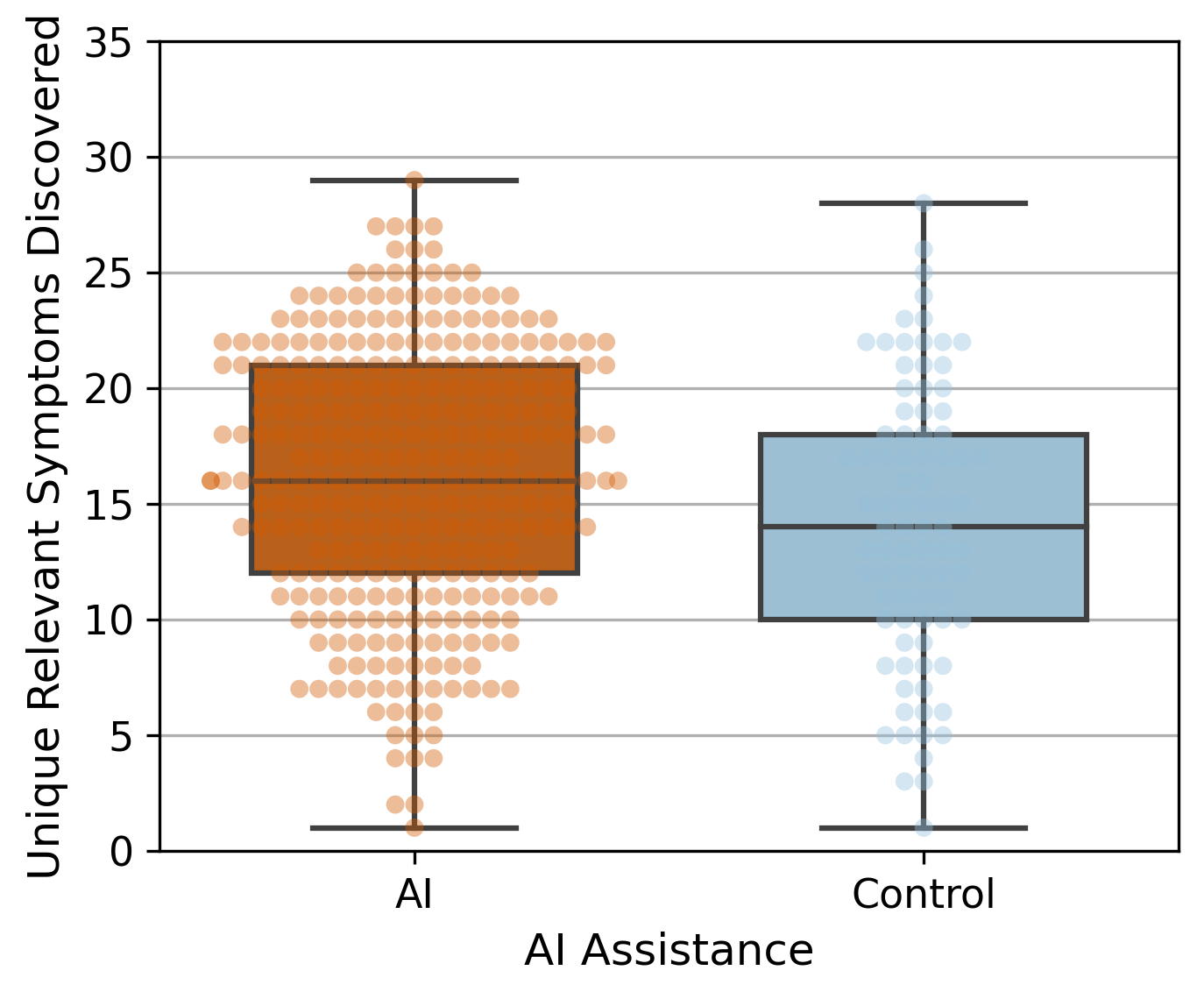}
    \caption{The spread of relevant, unique symptoms discovered between participants who did and did not receive \ai\ guidance. We observed that participants who received \ai\ guidance found more unique symptoms related to the epidemic than the \ctrl.}
    \label{fig:keyword_diversity_e_vs_hard}
    \vspace{-2.2em}
\end{figure}

\noindent
\textbf{Does \ai\ guidance encourage symptom diversity?}\hspace{1em}
Figure~\ref{fig:keyword_diversity_e_vs_hard} shows the spread of relevant symptoms discovered between participants who did and did not receive \ai\ guidance. We conducted an independent Mann-Whitney U test, which is robust to unequal sample size, to test whether there was a difference in ~\textsc{symptom diversity} between the \ctrl\ ($median = 14.0$, IQR $(10.0, 18.0)$) and \ai-guided ($median = 16.0$, IQR $(12.0, 21.0)$) groups. We found that the \ctrl\ group \textit{interacted with a significantly less diverse set of relevant symptoms} than the \ai\ guided group $(U = 10477.5, p = .0054, \eta^{2} = .02)$.  

\noindent
\textbf{Does the presence of \ai\ guidance bias exploration?}\hspace{1em}
To gain a deeper understanding of whether \ai\ guidance biased data exploration, we analyzed how the participants who did and did not receive \ai\ guidance covered the dataset during their investigation by comparing the distribution of locations explored. See supplementary materials for visual distributions of locations investigated for the control and \ai\ assistance groups that completed the hard task. \textit{We observed similar exploration patterns between the two groups, with the \ctrl\ exploration being moderately more biased than the \ai-guided participants towards the lower right corner of the city.}

\noindent
\textbf{Is there a relationship between suggestion usage and symptom diversity?}\hspace{1em}
In the previous analysis, we observed that the \ai-guided participants interacted with a significantly more diverse set of symptoms. Based on this finding, we then explored whether there was a relationship between \usage\ and symptom diversity. A Spearman's rank correlation was computed, and we observed a weak positive correlation ($r$(307) $= .312, p < 0.001)$), \textit{supporting that as the participants utilized more suggestions, the participants were able to uncover more relevant symptoms}.

\section{Discussion}
This user study aimed to observe how task difficulty and the promotion of transparency impact users' trust, \ai\ suggestion usage, and data exploration whilst performing an analytical data foraging task with \ai\ guidance. We discuss the implications of our findings.

\subsection{Users were more likely to use suggestions for hard tasks.}

Our analysis of how users interact with \ai\ suggestions considered task difficulty as a variable. In particular, we examined a scenario where guidance might be more valuable but prone to errors. We observed that users tend to rely on \ai\ suggestions more when finding relevant data is more challenging. While this is not surprising, our findings reveal that, at times, users may depend too much on suggestions. In particular, our study found that participants in more challenging conditions were more likely to accept suggestions from \ai\ that were not entirely accurate or appropriate for the task. Participants who received no transparency similarly overrelied on \ai\ suggestions compared to those who received some level of transparency. Moreover, we corroborate the prior findings in the \ai\ community~\cite{vasconcelos2022explanations}, which showed an overreliance on \ai\ suggestions in more difficult tasks and demonstrate that this overreliance extends to data exploration with \va\ systems.

Designers should consider these findings that indicate a higher dependence on \ai\ for challenging tasks when developing \va\ tools with \ai\ guidance. One solution to account for higher dependence is to create tools that can automatically detect the difficulty level of a task and adjust the level of \ai\ support accordingly. For instance, if a task is deemed easy, the tool can reduce the frequency or intrusiveness of \ai\ suggestions but provide more guidance for challenging tasks. One way to determine task difficulty is by analyzing the user's interaction with the system. The \ai\ can learn from the duration of completing sub-tasks, the number of repeated or redundant interactions, and the frequency of guidance requests. These methods could rely on the existing body of work on analytic provenance~\cite{xu2020survey}. Additionally, designers could consider incorporating user feedback, such as the user's belief in their domain expertise or the difficulty of the task.


\subsection{Trust remained unaffected by transparency.}

Our study aimed to investigate the lack of trust and suggestion usage in some participants, which was observed in Monadjemi et al.~\cite{monadjemi2022guided} with their \va\ scenario. We incorporated various levels of transparency to help establish an appropriate level of trust. Our findings indicate that transparency did not significantly impact the user's interactions with the \ai\ suggestions or subjective trust. Moreover, we contradict previous research that suggested transparency can enhance users' trust, confidence, and understanding of \ai\ systems~\cite{bansal_does_2021,bucinca_proxy_2020}. 

\va\ may have a unique advantage in terms of the level of trust that users have in the system. If the data source is reliable, analysts may tend to approach problems with a belief in transparency and agency, which are common antecedents of trust. Therefore, the broader findings of \ai\ may not necessarily apply to \va. Additionally, we observed null results which could be due to a ceiling effect. Since our participants' baseline trust was already high, the impact of transparency may not be significant.

It is also plausible that the lack of significant results was due to the simplicity of the transparency techniques employed. To provide more context for the user's final decision, it may be helpful to supplement the confidence value with additional explanations. However, this requires further investigation to determine the appropriate balance of awareness, to avoid information overload, and unwanted interaction behaviors such as aversion or overreliance.



\subsection{A possible link between usage and exploration diversity.}
\textcolor{edits}{A potential risk of providing \ai\ guidance within \va\ systems is introducing or reinforcing the analyst's bias~\cite{wall_four_2018}.} Our study found no evidence of bias, as suggestion usage led participants to discover a more diverse set of symptoms. However, this may highlight a limitation of this study as this phenomenon could be explained by the ``wisdom of the crowd effect.'' The \textit{wisdom of the crowd} refers to the collective intelligence of a group of individuals, which can sometimes lead to better outcomes than those made by individual experts. In contrast, many real-world \va\ scenarios do not benefit from such group dynamics. Future work (e.g., utilizing a case study approach or priming to elicit cognitive biases) could shed more light on the risks associated with designing and implementing \ai-guided \va\ systems.

\vspace{-0.5em}
\section{Limitations and Future Work}
\textcolor{edits}{To balance control and generalizability, we designed this study using a \textsc{vast} Challenge task that mimics real-world analytic tasks. In this task, our participants explored a dataset related to an epidemic using a map-based visualization, which means that our results may not apply to other scenarios beyond the specific task used. However, we observed that some of our findings confirm prior research. For example, we found that the use of \ai\ can improve data discovery~\cite{monadjemi2022guided}, and participants are more likely to accept \ai\ suggestions when the task is challenging or the model is uncertain~\cite{wang_are_2021, jacobs_how_2021}. Considering that map-based visualizations are common on the web~\cite{battle2018beagle}, and the study's exploratory nature allowed participants to adopt their desired exploration strategies, we believe that our findings can be widely applicable.}

\textcolor{edits}{Our system leveraged the user's analytical provenance to continually retrain the model based on updated observations to suggest relevant data points. Since the \ai\ did not have access to the ground truth beyond labels provided by the user, the suggestions were contingent upon the data tagged for investigation. Given the relatively high average \ai\ accuracy in both the \hard\ and \easy\ task conditions as seen in Figure~\ref{fig:sugg_acc_rel}, we can see the potential of utilizing \ai\ suggestions for collaborations in exploration and decision-making settings. Especially, there is an opportunity for future work to explore how transparency can affect subjective trust and suggestion usage in more complex and nuanced \ai-guided \va\ scenarios.}

Unlike previous work~\cite{zhangEffectConfidenceExplanation2020, linderHowLevelExplanation2021, bansal_does_2021,bucinca_trust_2021}, we observed no effect of transparency on trust and suggestion usage. The transparency levels tested in this study do not represent all the established \textsc{xai}\ methods in \ai-assisted decision-making. We chose these transparency techniques based on what would be best suited for our specific \ai\ algorithm and \va\ scenario. Therefore, further research is necessary to explore the observed high baseline trust phenomenon with the \va\ system, the design of explanations, and transparency-promoting techniques for visualization interfaces. 

Although our crowdsourced approach captured a large and diverse study population, we acknowledge that our controlled user study has some limitations and confounds. One limitation is that the participants recruited for the study were neither asked nor screened about their experiences with visual data exploration tasks or their familiarity with \ai\ guidance before participating. 
Therefore, our findings may not apply to expert user scenarios. Moreover, our participants could have altered their behavior or responses due to their awareness of being studied. This issue may have impacted our observations, limiting our ability to draw accurate conclusions about users' natural behavior. Also, our between-subjects comparisons prevented us from controlling for individual differences, potentially limiting our ability to capture the full extent of how the different transparency levels impact users' trust and interactions with the \ai.

Lastly, consistent with prior work~\cite{monadjemi2022guided}, we presented 10 relevant suggestions at a time from the \ai\ to the participants who received guidance. 
Future work is needed to explore how varying the number of suggestions displayed at once impacts suggestion usage, data exploration, and cognitive load.

\vspace{-0.5em}
\section{Conclusion}
This paper explored the impact of \ai\ suggestions on user behavior during data exploration, specifically with an \ai-guided \va\ tool. We aimed to understand how task difficulty and transparency of the \ai\ can affect users' trust, interactions, and data exploration. Our results suggest that participants tended to trust the \ai\ regardless of the amount of transparency provided. We observed that the more difficult the task, the more likely users were to rely on suggestions provided, despite the \ai\ providing suggestions at a lower rate of accuracy. Furthermore, we demonstrate that the level of detail provided to promote transparency had no measurable impact on data exploration, suggestion usage, and trust. Finally, we discuss the implications of our results, including some of the promises and challenges of promoting transparency of the \ai\ in guided data discovery tools. Our findings underscore the importance of transparency in such \ai-guided data discovery tools, prompting further inquiry into its role in fostering appropriate reliance on \ai\ within \va\ systems.



\section*{Acknowledgements}
This work is supported in part by the National Science Foundation under Grant No. OAC-2118201 and IIS-2142977.

\bibliographystyle{eg-alpha-doi}
\bibliography{main}




\end{document}